\documentclass[a4paper,11pt]{article}
\usepackage{pos}
\usepackage{physics}
\usepackage{float}
\usepackage{subcaption}
\usepackage{amsmath}
\usepackage{dsfont}
\usepackage{tikz}
\usepackage{tabularx}
\usepackage{tikz-cd}
\usepackage{MnSymbol}
\usepackage{graphicx}
\usetikzlibrary{decorations.markings}

\title{Phases at finite winding number of an Abelian lattice gauge theory}

\author*[a,b]{Paolo Stornati}
\author[c]{Debasish Banerjee}
\author[a]{Karl Jansen}
\author[d]{Philipp Krah}

\affiliation[a]{Deutsches Elektronen-Synchrotron DESY,  Platanenallee 6, 15738 Zeuthen}

\affiliation[b]{ICFO, The Barcelona Institute of Science and Technology, Av. Carl Friedrich Gauss 3, 08860 Castelldefels (Barcelona), Spain
}

\affiliation[c]{Theory Division, Saha Institute of Nuclear Physics, HBNI, 1/AF Bidhan Nagar, Kolkata 700064, India}

\affiliation[d]{TU Berlin,
    Institute of Mathematics,
    Straße des 17. Juni 136, 10623 Berlin, Germany}

\emailAdd{paolo.stornati@icfo.eu, debasish.banerjee@saha.ac.in, karl.jansen@desy.de, krah@math.tu-berlin.de}

\abstract{Pure gauge theories are rather different from theories with 
pure scalar and fermionic matter, especially in terms of the nature of 
excitations. For example, in scalar and fermionic theories, one can 
create ultra-local excitations. For a gauge theory, such excitations 
need to be closed loops that do not violate gauge invariance. In this 
proceedings, we present a study on the condensation phenomenon associated 
with the string-like excitations of an Abelian lattice gauge theory. These 
phenomena are studied through numerical simulations of a $U(1)$ quantum 
link model in 2+1 dimensions in a ladder geometry using matrix product 
states. In this proceedings, we show the existence of ground states 
characterized by the presence of such string-like excitations. These 
are caused due to the condensation of torelons. We also study the 
relationship between the properties of the plaquettes in the ground 
state and the presence of such condensation phenomenon.}

\FullConference{%
 The 38th International Symposium on Lattice Field Theory, LATTICE2021
  26th-30th July, 2021
  Zoom/Gather@Massachusetts Institute of Technology
}

\begin{document}
\maketitle

\section{Introduction}
Quantum simulators are promising concepts, which enable experimental 
realizations of theoretical models. They are increasingly used to study 
the physics of lattice quantum field theories\cite{Ba_uls_2020}. Nevertheless, 
classical simulations are still the most reliable method to study lattice 
field theory, with results close to the thermodynamic limit. In particular, 
tensor networks are a great tool to study lattice field theory in situations 
where standard Markov chain Monte Carlo is known to fail \cite{Ba_uls_2013}. 
In addition, such classical tensor network simulation  methods can guide and 
benchmark quantum simulations since they share the same Hamiltonian. In this 
regard, such classical simulation methods are very well positioned to guide 
frontier research in this field.

  In this work, we expand on exploring the phases which can be realized in a 
strongly interacting lattice gauge theory. It is well-known that for a pure 
gauge theory, it is possible to have loop operators which wind around the 
spatial volume, which we will denote as {\em winding number operator} and 
which are termed as torelons \cite{Koller:1987fq}. For the non-Abelian gauge 
groups $SU(2)$ or $SU(3)$ commonly used for studies of confinement, these 
operators are associated with a $Z(2)$ or $Z(3)$ center symmetry. In the case 
of the Abelian $U(1)$ lattice gauge theory, however, such operators carry a 
global $U(1)$ quantum number. Consequently it becomes possible to couple a 
chemical potential to this winding number operator. Here, we study the 
condensation phenomena of the string-like excitations (expectation value of 
the winding number operators) at large values of the chemical potential coupled to the winding number operator. 
  
  The presence of string-like excitations that can spread over the entire 
lattice extent in a cylindrical geometry is of great interest \cite{Michael1987, Zach1998}.
In particular, one can ask to what extent the confining properties of the 
ground state are disturbed when such excitations are present. To illustrate 
why this can be the case, consider the physics of an interacting bosonic or 
fermionic model with a global symmetry. A pedagogical example is the XY-model, 
where one can also couple a chemical potential. In the absence of the chemical 
potential at weak couplings, the excitations are short ranged, and the system 
has a mass gap. However, upon subjecting the system to a sufficiently large 
chemical potential, the particle excitations condense, and long-range correlations 
(signalled by a non-zero value of the superfluid stiffness) appear in the system
\cite{DB2010}. We aim to investigate such physics in the pure gauge theory context. 
Furthermore, these string-like excitations of a gauge theory have their own unique 
dynamics, which can in principle be very different from the dynamics of point-particles 
in scalar theories or mesons in fermionic theories. 
  
  The goal of this study is to demonstrate the presence of the string excitations 
in the ground state numerically. In scalar or fermionic theories, the expectation 
value of the particle number operator presents discrete jumps when the chemical 
potential is increased (e.g., \cite{DB2010, PhysRevLett_118_071601}). When we raise 
the system’s density, a higher particle state may have lower energy than the corresponding 
lower particle state. We expect the same behaviour for string-like excitations in our 
Abelian gauge theory. This sting-like excitations divide the Hilbert space in sectors 
that are exponentially hard to explore by Quantum Monte Carlo simulations. For this reason,
tensor network algorithms have been chosen to simulate this model. 

\section{Quantum link model}
While a Wilson-type Abelian gauge theory has an infinite-dimensional Hilbert space for 
every link, the quantum link models (QLMs) regulate this infinite-dimensional Hilbert space 
in a completely gauge-invariant fashion \cite{HORN1981149,ORLAND1990647,Chandrasekharan_1997}.
This is achieved by replacing the quantum rotors in the Wilson theory for a quantum spin
in the QLM, such that the resulting finite-dimensional Hilbert spaces have sizes $(2S + 1)$ 
locally on the links. The spin representations take the values $S = 1/2, 1, 3/2, \dots$ and 
the Wilsonian theory is obtained when $S \rightarrow \infty $ \cite{PhysRevD_63_085007}. In 
this work we consider the $U(1)$ quantum link model in 2+1 dimensions with the spin $S=1/2$ 
representation. Due to the low finite dimensional Hilbert space of this model, quantum simulator
proposals have been made \cite{2014Marcos, 2020Celi}, or this could be simulated on larger 
NISQ devices \cite{huffman2021realtime}.

The Hamiltonian of the system is defined as:
\begin{equation}\label{eq:hamiltonian}
    \mathcal H = - J \sum_\square (U_\square + U^\dag_\square ) + \sum_\square \lambda (U_\square + U^\dag_\square )^2\,,
\end{equation}
where we have defined the plaquette operator $U_\square= U_{x,\mu}U_{x+ \hat \mu, \nu}U^\dag_{x+\hat \nu,\mu}U^\dag_{x,\nu}$
and the operators $U_{x,\mu}$ act on the links. The Hamiltonian has a local $U(1)$ gauge symmetry 
and the generator of the symmetry is: 
\begin{equation}
G_x=\sum_\mu\left(E_{x-\hat \mu,\mu}-E_{x,\mu} \right)=\sum_\mu\left(S^z_{x-\hat \mu,\mu}-S^z_{x,\mu} \right)
\end{equation}
where $E_{x,\mu}$ is the electric field operator. The gauge-field operator is canonically conjugate
to the electric field operator, i.e., $[E_{x,\mu}, U_{y,\nu} ]=U_{x,\mu} \delta_{\mu,\nu} \delta_{x,y}$ and 
$[E_{x,\mu}, U^\dag_{y,\nu} ]= -U^\dag_{x,\mu} \delta_{\mu,\nu} \delta_{x,y}$. As explained before, we can
use quantum spins as degrees of freedom, which satisfy the above commutation relations. In the $S^z$ basis, 
we can represent: $U_{x,\mu}=S^+_{x,\mu}$, $U^\dag_{x,\mu}=S^-_{x,\mu}$ and $E_{x,\mu}=S^z_{x,\mu}$, with 
the spin raising and lowering operators $S^+$ and $S^-$.

\tikzset{middlearrow/.style={
        decoration={markings,
            mark= at position 0.65 with {\arrow[scale=2]{#1}} ,
        },
        postaction={decorate}
    }
}
\begin{figure}[t!]
\centering
\resizebox{0.7\textwidth}{!}{  
\begin{tikzpicture}
\def \dx{2};
\def \dy{2};
\def \nbx{8};
\def \nby{3};
\foreach \x in {1,...,\nbx} {\foreach \y in {1,...,\nby} {\ifthenelse{\x=2}{}{\node at (\x*\dx,\y*\dy) [circle, fill=black] {};}}}
\draw (1*\dx,\dy) -- ( 1*\dx,\nby*\dy);
\foreach \x in {3,...,\nbx} {
        \draw (\x*\dx,\dy) -- ( \x*\dx,\nby*\dy);  
}
\foreach \y in {1,...,2} {
    \draw [thick](3*\dx,\y*\dy) -- ( \nbx*\dx,\y*\dy);
    \draw [thick](\dx,\y*\dy) -- (\dx*1.7,\y*\dy);
}

\foreach \y in {3} {
    \draw [thick, dashed](3*\dx,\y*\dy) -- ( \nbx*\dx,\y*\dy);
    \draw [thick, dashed](\dx,\y*\dy) -- ( \dx*1.7,\y*\dy);
}

\draw[->] (1,1 ) -- (+\dx,1) node [pos=0.66,below] {$\hat{x}$};
\draw[->](1,1 )  -- (1, +\dy) node [pos=0.66,left] {$\hat{y}$};
\draw[->](1,1 )  -- (1, +\dy) node [pos=0.66,left] {$\hat{y}$};
\node[text width=3cm,above] at (2*\dx,1*\dy) {$U_{x,\mu}$};
\node[left] at (1*\dx,1.5*\dy) {$U_{x,\nu}$};

\node[thick] at (4.5*\dx,1.5*\dy) {\Huge$\circlearrowleft$};
\def \xO{8};
\def \yO{2};
\draw[thick,middlearrow={latex},draw opacity=0] (\xO,\yO) -- ( \xO+\dx,\yO);
\draw[thick,middlearrow={latex},draw opacity=0] (\xO+\dx,\yO) -- (\xO+\dx,\yO+\dy);
\draw[thick,middlearrow={latex},draw opacity=0] (\xO+\dx,\yO+\dy) --( \xO,\yO+\dy);
\draw[thick,middlearrow={latex},draw opacity=0] (\xO,\yO+\dy) --( \xO,\yO);
\node[thick] at (6.5*\dx,2.5*\dy) {\Huge$\circlearrowright$};
\def \xO{6*\dx};
\def \yO{2*\dy};
\draw[thick,middlearrow={latex},draw opacity=0] ( \xO+\dx,\yO)--(\xO,\yO);
\draw[thick,middlearrow={latex},draw opacity=0] (\xO+\dx,\yO+\dy) -- (\xO+\dx,\yO);
\draw[thick,middlearrow={latex},draw opacity=0] (\xO,\yO+\dy) --(\xO+\dx,\yO+\dy);
\draw[thick,middlearrow={latex},draw opacity=0] (\xO,\yO) --( \xO,\yO+\dy);
\node[thick] at (6.5*\dx,1.5*\dy) {\Huge$\ncirclearrowright$};
\def \xO{6*\dx};
\def \yO{1*\dy};
\draw[thick,middlearrow={latex},draw opacity=0] ( \xO,\yO)--(\xO+\dx,\yO);
\draw[thick,middlearrow={latex},draw opacity=0] (\xO+\dx,\yO+\dy) -- (\xO+\dx,\yO);
\draw[thick,middlearrow={latex},draw opacity=0] (\xO,\yO) --( \xO,\yO+\dy);
\end{tikzpicture}
}


 \caption{Ladder geometry of the lattice and its local basis used inside the DMRG algorithm. 
The periodicity in $\hat{y}$ is indicated by the dashed lines. Furthermore we show the two 
flippable plaquettes ($\circlearrowleft$, $\circlearrowright$) and a non flippable plaquette 
($\ncirclearrowright$). }
  \label{fig:lattice_conf}
\end{figure}
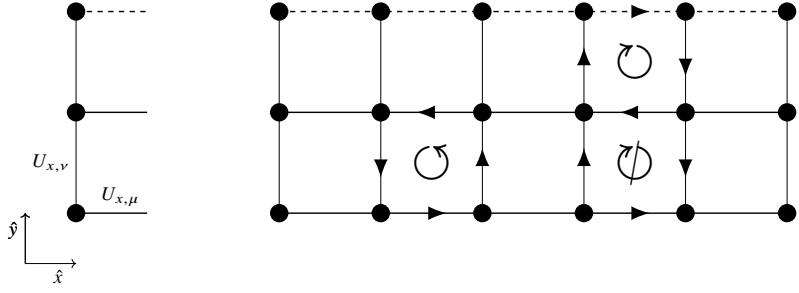

\section{Symmetries of the model}
We consider a 2+1 dimensional lattice with extension $L_x\times L_y $, shown in 
\autoref{fig:lattice_conf}. In addition to the $U(1)$ gauge symmetry, the Hamiltonian also 
has the point group symmetries (translation, (discrete) rotation, and parity). The 
Hamiltonian is invariant under the charge conjugation symmetry ($Z_2$). Moreover, the system 
has a global invariance with respect to winding number transformations ($U(1)\otimes U(1)$). 
The generators of these symmetries are the winding number operators: 
\begin{eqnarray}
W_x = \frac{1}{L_y}\sum_{y}^{L_y} S^z_{(x,y),\hat x}\quad \text{and}\quad
W_y= \frac{1}{L_x}\sum_{x}^{L_x} S^z_{(x,y),\hat y}\,,
\end{eqnarray}
respectively for the $x$ direction and for the $y$ direction. The generators of this symmetry 
commute with the Hamiltonian $\left[H \, ,W_x \right]=\left[H \, ,W_y \right]=0 $ and are thus simultaneously diagonalizable. We therefore add these operators to the Hamiltonian with the coefficient $\mu_x$ and $\mu_y$ used as 
chemical potential in the $x$ and in the $y$ direction. The resulting Hamiltonian is:
\begin{equation}\label{eq:hamiltonian}
    H = - J \sum_\square (U_\square + U^\dag_\square ) + \lambda \sum_\square (U_\square + U^\dag_\square ) ^2 + \mu_x \sum_x W_x + \mu_y \sum_y W_y\,.
\end{equation}
The operators $W_x$ and 
$W_y$ allow for the formation of stringy excitations that spread over the lattice. Moreover, 
in analogy with the particle number operators in fermionic and scalar theories, the ground state 
energy is given by:
\begin{equation}
E_{GS}= E_ \mathcal H - \mu_x N^x - \mu_y N^y\,,
\end{equation}
where $N^x$ and $N^y$ are the expectation  values of $ \sum_x W_x$ and $\sum_y W_y$. We now consider 
all the possible states one plaquette can have. Since a plaquette is formed, in our quantum link model, 
by four different spins 1/2, the number of states is $n_{\mathrm{states}}=2^4=16$. The two plaquette 
operators present in the Hamiltonian are:
 \begin{eqnarray}\label{eq:plq_operator_hamiltonian}
U_\square= S^+_{x,\mu}S^+_{x+ \hat \mu, \nu}S^-_{x+\hat \nu,\mu}S^-_{x,\nu}\quad \text{and}\quad
U^\dag_\square= S^-_{x,\mu}S^-_{x+ \hat \mu, \nu}S^+_{x+\hat \nu,\mu}S^+_{x,\nu}\,.
\end{eqnarray}
In our representation, a plaquette can be any combination of the four different spins that compose 
the plaquette (e.g., $\ket{\downarrow \downarrow \downarrow \uparrow}$). 
Between all the possible 16 states a plaquette can have, only two are not annihilated by the action 
of the $U_\square$ and $U_\square^\dag$ operators. The two possible states $\ket{\circlearrowright}$ 
and $\ket{\circlearrowleft}$ are visualized in \autoref{fig:lattice_conf}. We define $\ket{\circlearrowright}$ 
as the classical state in which the plaquette state has a clockwise orientation 
($\ket{\uparrow \uparrow \downarrow \downarrow }$).  
We define $\ket{\circlearrowleft}$ as the classical state in which the plaquette state has a counter-clockwise 
orientation ($\ket{\downarrow \downarrow \uparrow \uparrow}$).  We define the flippability operator 
$O_{\mathrm{flipp}}$ as the difference between the fippability in even and odd sites in the x direction 
over the lattice: 
\begin{equation}\label{eq:delta_wind_relation}
O_{\mathrm{flipp}} =  \sum_\square (-1)^x(U_\square+ U^\dag_\square )^2 =  \sum_\square (-1)^x(U_\square U^\dag_\square + U^\dag_\square U_\square )\,.
\end{equation}
We note that, within the quantum link formulation, the operator $U_\square U^\dag_\square$ is not the 
identity operator since $U_\square$ is not unitary. Hence, this operator can measure non-trivial correlations.  
In the numerical studies, we will demonstrate a correlation between this operator and the number of excitations 
present in the ground state.
\begin{figure}[t!]
\centering
\includegraphics[width=0.8\textwidth]{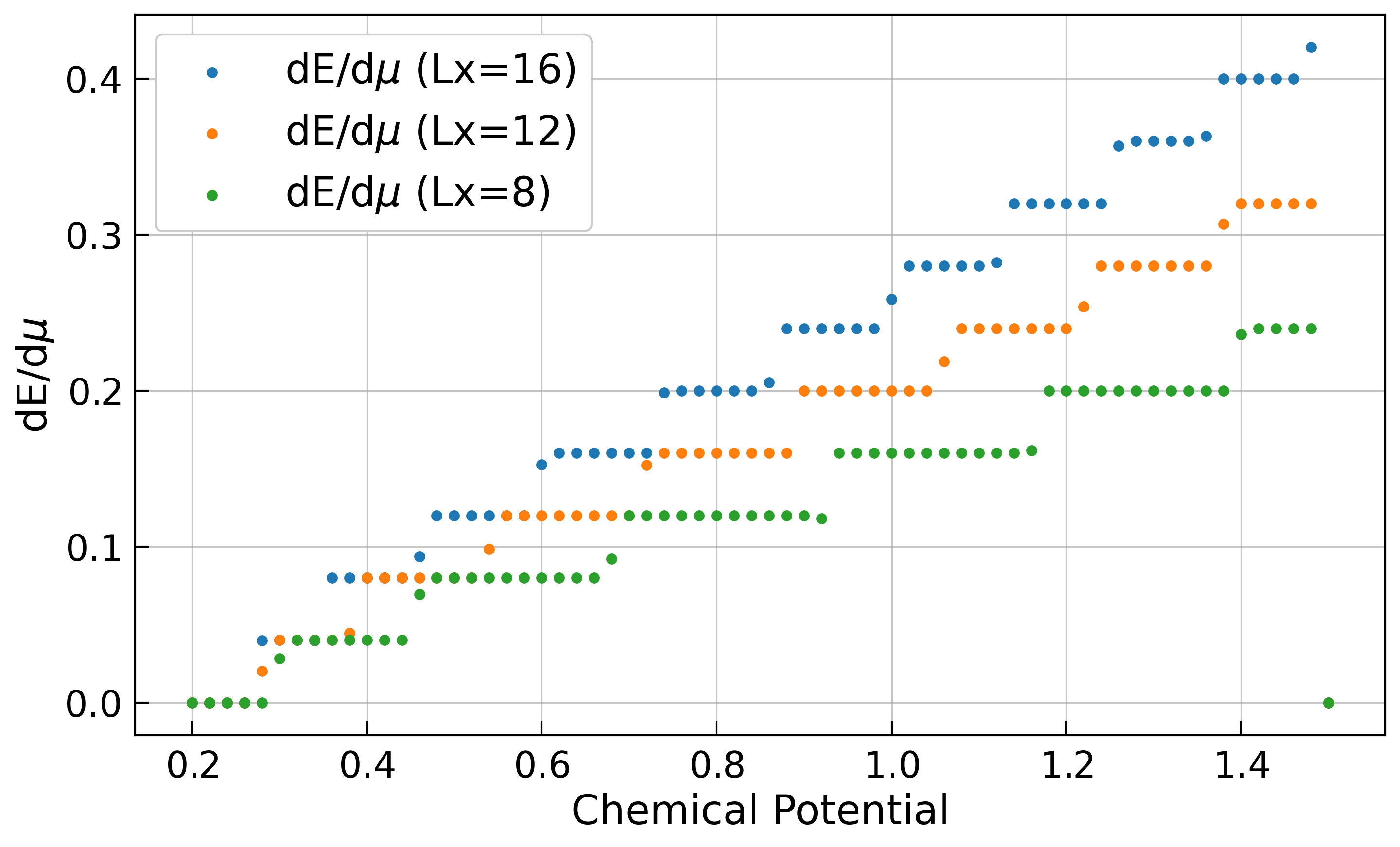}
\caption{ Derivative of the energy as a function of the chemical potential.
}
\label{fig:winding_numbers}
\end{figure}
\section{Numerical simulations}
In our numerical studies, the wave function of the system is represented by matrix product states 
and the ground state of the model is computed with the Density matrix renormalization group algorithm 
(DMRG)\cite{white_dmrg}. The DMRG algorithm performs extremely well in 1+1 dimensions. For this reason, 
we have studied the system in a ladder geometry (see \autoref{fig:lattice_conf}) with open boundary 
conditions in the $x$ direction and periodic boundary condition in the $y$ direction (the theory lives 
on a cylinder). The local basis used inside the DMRG algorithm is shown in \autoref{fig:lattice_conf}. 
It is staged together as a chain to form the ladder geometry of our lattice. The dashed lines indicate 
the links on the  periodic boundary. In the numerical studies, We keep $\mu_x=0$ in order to prevent the system from generating a winding in the x-direction, which would otherwise kill all dynamics.

The commutator $[H,W_y]$ is no longer zero when open boundary conditions are imposed. However, the 
non-commuting terms only appear in the boundaries, and therefore for large enough extents in the 
x-direction this does not pose a problem. Moreover, note that the effect of the $W_y$ operator can 
also be realized by imagining a static charge-anti-charge pair at the boundaries, which inject a 
background electric flux into the system. In our numerical simulations, we explore the effect of 
the chemical potential ($\mu_y$) on the ground state properties of the theory. We fix the parameter 
in the Hamiltonian to be $J=1$ and $\lambda=-1$. We study the properties of the ground state as a 
function of the chemical potential $\mu_y$. 

\begin{figure}[t!]
\centering
\includegraphics[width=0.8\textwidth]{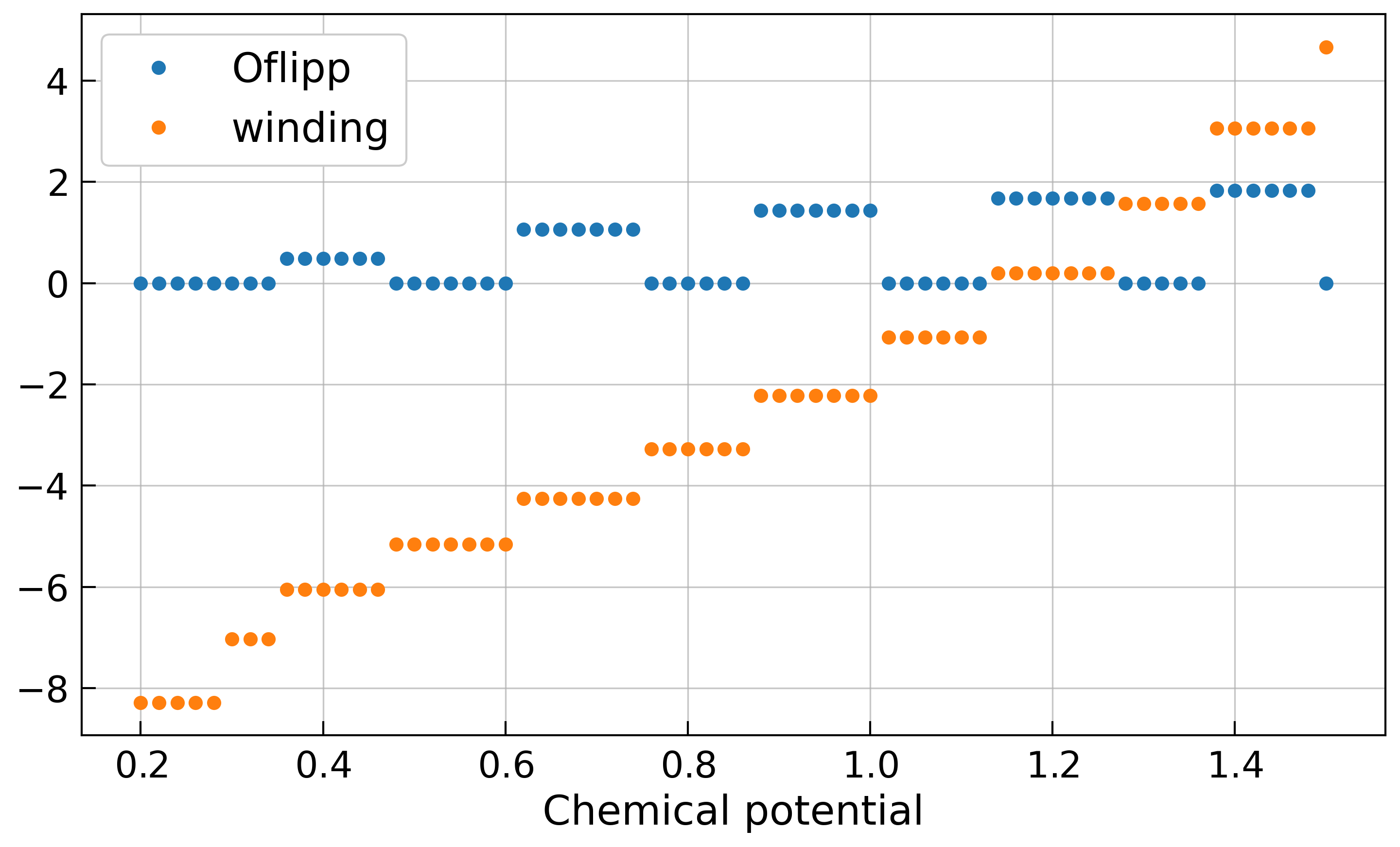}
\caption{ Correlation between flippability and winding sectors.
}
\label{fig:correlation}
\end{figure}

In \autoref{fig:winding_numbers} we show the numerical derivative of the ground state energy 
extrapolated in the infinite bond dimension limit with respect to the chemical potential as a function 
of the chemical potential. The derivative of the expectation value of the energy of the ground state 
corresponds to the expectation value of the winding number operator on the ground state. The winding 
number operator can be seen as an operator that counts the number of string-like excitations in the 
ground state. In this case the excitations are strings that propagate through the whole lattice. 
Discrete jumps in the number of excitations in the ground state are clearly visible in \autoref{fig:correlation}. 
When the chemical potential is increased, the system favors a lower energy state with more particles 
in the ground state.

In \autoref{fig:correlation} we plot two different quantities as a function of the chemical potential. 
In blue we plot the absolute value of the flippability of the state defined in \autoref{eq:delta_wind_relation}.
When the values in the plot are close to zero, they are zero to numerical precision ($ 10^{-10}$ in our case). 
We plot the absolute value of the expectation value because the operator is symmetric under charge conjugations. 
In orange, we plot the expectation value of the winding number operator. It is clear from this plot that the 
two quantities are correlated. Starting from small chemical potential, we have a low number of excitations and 
a zero flippability. When we increase the chemical potential, we see a transition in the winding number and in 
the flippability. At the next transition point for the winding number operator, the flippability goes back to 
zero. This pattern is repeated when the chemical potential is increased. Since the flippability operator is 
defined as the difference of the flippability of the single plaquettes in even and odd site, we can measure 
if we have an even or odd number of excitations in the ground state by computing the expectation value of 
the flippability.

\section{Conclusions}

In this proceeding, we analyzed a $U(1)$ quantum link model in 2+1 dimensions in a ladder geometry. Having 
motivated the importance of coupling a chemical potential to the winding number operators, we have studied 
the properties of the ground state of the model at finite volume and increasing values of the chemical 
potential. We have demonstrated how the increase of chemical potential corresponds to the condensation 
of string excitations in the ground state. In fact, increasing the chemical potential changes the 
flippability operator, as well as the number of string-like excitations present in the ground state in 
discrete steps. In bosonic and fermionic theories, increasing the chemical potential causes non-trivial 
particle numbers states to be the ground state. Similarly, in this case,  when we increase the chemical 
potential, we find the ground state to  comprise non-zero winding numbers. This causes the expectation 
value of the winding numbers to change in discrete steps. We have also seen a correlation between the 
winding number sectors and the average flippability of the plaquettes. With the finite dimensional Hilbert 
space at each link, this model is also a good candidate for cold atom simulations without resorting to any 
further truncation \cite{Lewenstein_2007}.

\acknowledgments
P.S. acknowledges support from Agencia Estatal de Investigación (“Severo Ochoa” Center of Excellence CEX2019-000910-S, Plan National FIDEUA PID2019-106901GB-I00/10.13039 / 501100011033, FPI) ),, Fundació Privada Cellex, Fundació Mir-Puig, and from Generalitat de Catalunya (AGAUR Grant No. 2017 SGR 1341, CERCA program).

P.K. acknowledges support from the Research Training Group ”Differential Equation- and Data-driven Models in Life Sciences and Fluid Dynamics: An Interdisciplinary Research Training Group (DAEDALUS)” (GRK 2433) funded by the German Research Foundation (DFG).

\bibliographystyle{JHEP}
\bibliography{publications}

\end{document}